\documentclass[reprint,showpacs,citeautoscript,secnumarabic,longbibliography,nobibnotes, amsmath,amssymb,aps,prb,floatfix
]{revtex4-2}

\usepackage{graphicx}
\usepackage{dcolumn}
\usepackage{bm}
\usepackage{hyperref}
\usepackage{multirow}
\usepackage{soul}
\usepackage{color}
\usepackage{xspace}

\newcommand{\BiSe}{Bi$_2$Se$_3$\xspace}
\newcommand{\BiTe}{Bi$_2$Te$_3$\xspace}

\newcommand{\GM}{$\bar{\Gamma}-\bar{M}$\xspace}
\newcommand{\KTH}{KTH Royal Institute of Technology, Applied Physics, Materials and Nano Physics, AlbaNova universitetscentrum, 106 91 Stockholm, Sweden}

\hypersetup{colorlinks=true, linkcolor=blue, citecolor=blue, urlcolor=blue, unicode=false}

\begin{document}

\preprint{APS/123-QED}

\title{Accessing the spin structure of buried electronic states}

\author{M.~H.~Berntsen}
\email{mhbe@kth.se}
\affiliation{\KTH}
\author{O.~G\"{o}tberg}
\affiliation{\KTH}
\author{B.~M.~Wojek}
\affiliation{\KTH}
\author{O.~Tjernberg}
\affiliation{\KTH}

\date{\today}

\begin{abstract}
In spin- and angle-resolved photoemission spectroscopy (SARPES) the energy-momentum dispersion of electronic states in crystalline solids is measured along with the spin direction of the photoemitted electrons. The technique therefore allows for mapping out a material's band structure in a spin resolved fashion. By conducting SARPES measurements using low-energy photons, the spin sensitivity of the technique can be combined an increased bulk probe depth, provided by the large electron inelastic mean-free path at these kinetic energies, to directly access the spin structure of electronic states at buried interfaces. Here, we demonstrate this capability by using SARPES to determine the spin polarization of photoelectrons emitted from a 6-nm-thick film of the topological insulator Bi$_2$Se$_3$ using photons with an energy of 8.5 eV. By modelling the expected spin structure in the film, we show that the complex spin polarization that is observed is the integrated spin signal from spin-polarized states at the surface, bulk and buried interface (bottom surface) of the topological-insulator film. Our results therefore allows us to directly determine the spin texture of the buried Dirac interface state. This capability is highly attractive for state-of-the art spectroscopic measurements of the spin-physics at play in quantum-material based or spintronic devices where spin-polarized interface states define the operational principle of the devices.
\end{abstract}


\maketitle


The electronic structure of surfaces and interfaces in crystalline solids is an inherent part of a material's electronic properties~\cite{Heine1965,LuthBook}. Although the contribution from the boundaries of a specimen often can be neglected in bulk materials, electronic states located at surfaces or interfaces can display properties that are absent in the interior of the material and therefore become interesting from the point of view of creating tailored systems with specific electronic properties~\cite{Ohtomo2004,Zhao2016}. In particular, when the spatial dimensions of the sample are reduced, and the surface-to-bulk ratio increases, the functional electronic behavior of a system could become dominated by non-bulk electronic states. The importance of the physics of surfaces and interfaces for applications is clear from the impact it has had on shaping the operational principles, design and performance of semiconductor-based electronic devices in the past~\cite{Kroemer1983, Schulz1983, Grunthaner1986}.  

In recent years, the discovery of topological insulator materials with robust spin-polarized surface states~\cite{Zhang2009a,Chen2009,Xia2009}, along with a range of materials displaying large Rashba spin-split electronic states~\cite{Ast2007,Crepaldi2012,Berntsen2018,Sanchez2013}, has led to an increased interest in spin-physics and materials with electronic states that could be utilized for spintronic applications~\cite{Wolf2001}. Surfaces and interfaces are of particular interest also in spin-physics~\cite{Miron2010,Zhang2020,MohdYusoff2021} since breaking of the crystal inversion symmetry allows for the spin degeneracy to be lifted in the presence of a strong spin-orbit interaction. The ability to study and characterize not only the electronic, but also the spin, structure of buried interfaces is therefore becoming increasingly important.

One of the prime tools for experimentally determining a material’s electronic structure is angle-resolved photoemission spectroscopy (ARPES)~\cite{Sobota2021}. The technique measures the single-particle spectral function which provides access to information about the bare-band dispersion of the material as well as electron scattering rates and correlations in the system. By using a spin-sensitive electron analyzer one can map out the electronic band structure in a spin-resolved fashion and the technique is then referred to as spin-ARPES (SARPES)~\cite{Dil2009}.

When conducting (S)ARPES experiments one typically uses extreme ultraviolet (XUV) radiation with photon energies between 20~eV and 200~eV. At these energies, the corresponding inelastic mean-free path (IMFP) of photo-excited electrons in a solid is below 1~nm~\cite{Seah1979} and the technique is therefore highly surface sensitive. However, moving towards photon energies in the soft x-ray range increases the effective probe depth and enables measurements of the bulk electronic properties. The IMFP also increases quickly for kinetic energies below 20~eV and reaches approximately the same value (1-2~nm) at a kinetic energy of 10~eV as for 1~keV. This means that both the low and high photon energy ranges can be used to probe the electronic structure deeper into a material and thus give access to the electronic structure at buried interfaces. The low photon-energy range, however, has the added benefit of higher angular and energy resolutions, which is an advantage when studying the low-energy electronic spectrum in materials. This is particularly useful if one’s interest is in conductive properties of materials since this involves states close to the Fermi level. 

Accessing the spin- and electronic structure of a buried interface is, however, associated with some challenges. This is due to the fact that the photoemission signal from the interface mixes with the signal from both the bulk as well as the top surface of the material. In particular, when performing SARPES, the measured spin polarization can deviate significantly from the expected result from an isolated interface, bulk or surface state if spin polarized bands are also present in any of the other regions, respectively. Accounting for this mixing becomes an important step in the data analysis and emphasizes the need for having supporting theoretical models for comparison.

Here, we demonstrate the capability of low photon-energy SARPES to measure the spin structure of a buried interface state. The sample system used in our study is a 6-nm-thick film of the topological insulator (TI) \BiSe which has two-dimensional spin polarized Dirac states localized at both the vacuum-interface surface of the film as well as at the interface towards the Si-substrate on which it is grown. Our results show that the measured spin polarization displays a complex behavior which originates from the mixing of photoemission spin signals from the top and bottom Dirac states as well as a Rashba-split conduction-band quantum-well state (QWS) present in the bulk of the film. We present a simple phenomenological model of the spin structure in the sample that describes the experimentally determined spin polarization signal remarkably well, demonstrating how the different spin structures can be disentangled even in the case where mixing of the photoemission signals occurs. We believe this work will open up for future use of SARPES in retrieving the spin and electronic structure from buried electronics states, something that is of great value for research related to spintronic-device applications.

\section{Experimental}

\begin{figure*}
\includegraphics[width=.8\textwidth]{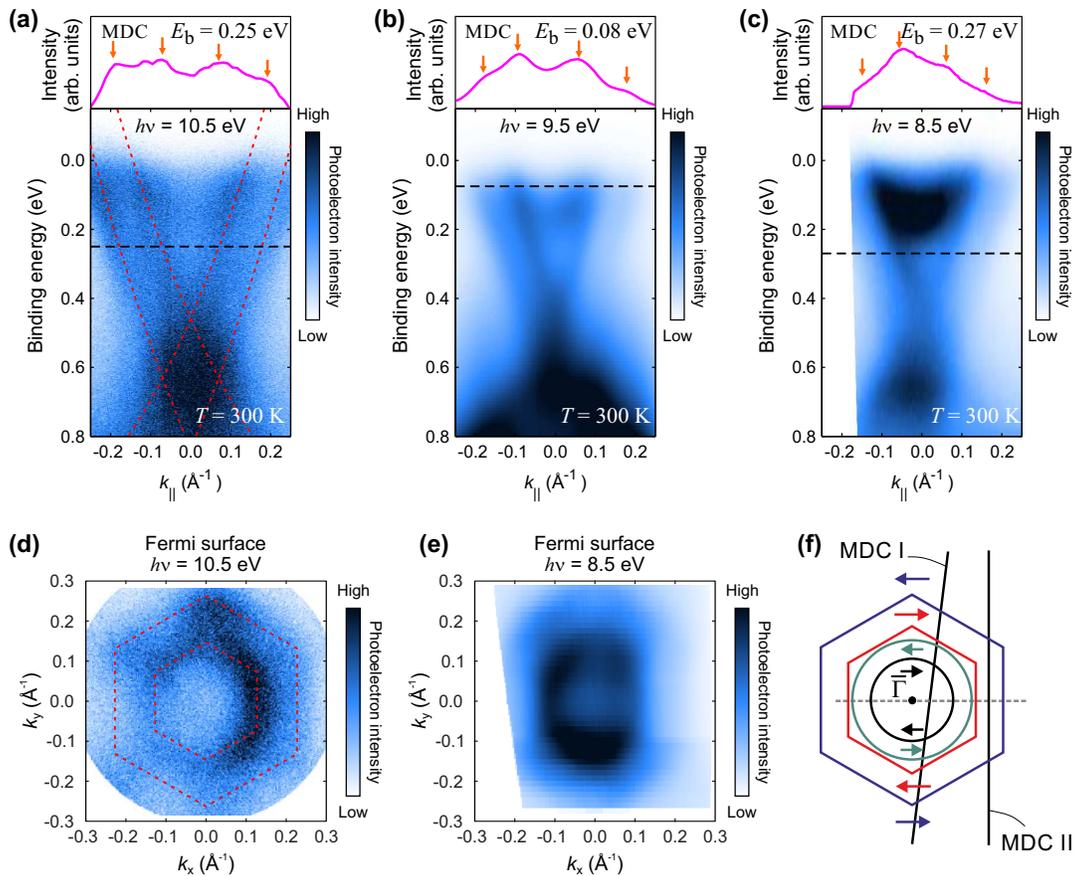}
\caption{\textbf{Photoemission intensity plots.} (a) ARPES spectrum of 6~QL \BiSe on Si(111) measured along \GM using the laser-ARPES setup ($h\nu = 10.5$~eV, $T = 300$~K). The dashed lines highlight the dispersion of the Dirac-like features in the spectrum. The horizontal dashed line indicates where the MDC plotted above the spectrum has been extracted. Arrows above the MDC indicate the peak position of the bands intersected by the MDC. (b) and (c) ARPES spectrum acquired at $T = 300$~K from the same sample as in (a) measured along \GM using the experimental setup in ref.~\onlinecite{Berntsen2010} with 9.5~eV and 8.5~eV photons, respectively. Dashed horizontal lines indicate positions of the plotted MDCs. (d) Experimental Fermi surface from the laser-ARPES measurement displaying two concentric hexagonal Fermi surfaces. (e) Fermi surface from the measurement in (c). (f) Schematic view of the Fermi surface consisting of a Rashba-split QWS (inner circles) in addition to the hexagonally warped Dirac-like states at the film surface and the interface to the substrate (inner and outer hexagons, respectively). The arrows indicate the spin polarization of the different states. The dashed line marks the position of the energy-momentum cut shown in (c); the solid lines marked with MDC~I and MDC~II indicate the paths along which the spin polarizations shown in Fig.~\ref{fig:mdcpol} are measured.}
\label{fig:arpes}
\end{figure*}

To demonstrate the capability of SARPES in determining the spin structure of a buried interface state we have studied thin films of the topological insulator (TI) \BiSe. Previous ARPES measurements on a similar sample have directly observed the Dirac states present at the top-surface of the film as well as at the interface between the TI and the Si(111) substrate~\cite{Berntsen2013}. The measurements in ref.~\onlinecite{Berntsen2013} were performed using low-energy photons (10.5~eV) which yielded an electron inelastic mean-free path (IMFP) in the range of 1-2~nm. The increased IMFP explains why the electronic structure of the interface Dirac state could be accessed even for a TI film with a thickness of approximately 6~nm (corresponding to 6 quintuple-layers (QL)~\cite{Zhang2009b}).

In the present study, we have studied another 6-QL-thick \BiSe film by SARPES using Mott polarimetry conducted at room temperature at the I3 beamline \cite{Berntsen2010} of the MAX-III synchrotron (MAX-lab, Lund University, Sweden). The thin-film sample was grown by co-evaporation of Bi and Se onto a Bi-terminated Si(111)-7$\times$7 substrate using electron-beam evaporation while keeping the substrate temperature at $270^\circ$C. The sample was grown in-situ at the BALTAZAR laser-ARPES facility~\cite{Berntsen2011} and characterized there using 10.5~eV photons, clearly observing the top-surface and buried interface Dirac states, as shown in Fig.~\ref{fig:arpes}(a) as a double X(V)-shaped feature. The inner X-feature is ascribed to the surface state and the outer (larger) V-shaped feature to the interface state. A constant energy cut taken at the Fermi level, shown in Fig.~\ref{fig:arpes}(d), clearly shows two hexagonally deformed Fermi surfaces from the two Dirac states. Before transporting the sample to the I3 beamline, a capping layer of Se was deposited on the sample. During transport to MAX-lab the sample was kept in a vacuum suit case under static vacuum conditions ($<10$~mbar). Once loaded into the preparation chamber of the SARPES end station, the sample was gently annealed to $220^\circ$C for 30 minutes in order to remove the Se-capping layer. After removal of the capping layer, the surface quality was checked with low-energy electron diffraction (LEED). A LEED pattern of similar quality as the one obtained from the freshly grown sample at the BALTAZAR facility was observed.

To further enhance the IMFP, the current study used photon energies of 9.5~eV and 8.5~eV, respectively, to collect energy-momentum cuts along $\bar{\Gamma}-\bar{M}$ of the surface Brillouin zone (SBZ) at a sample temperature of 300~K. The results are presented in Fig.~\ref{fig:arpes}(b) and (c) and display a double X(V)-shaped structure similar to the one seen in the laser-ARPES spectrum in Fig.~\ref{fig:arpes}(a). We notice that at low photon energies, the photoemission intensity in the different features present in the spectra varies rapidly with changed excitation energy. We therefore ascribe the intensity-ratio difference of the surface and interface state between panels Fig.~\ref{fig:arpes}(a)–(c) to such variations in photoemission intensity – most likely originating from matrix element effects in the photoemission process. Nevertheless, all three spectra show the presence of both the surface and interface Dirac states, more clearly visualized by the momentum distribution curves (MDCs) presented in each of the panels. The additional parabolic dispersing state present at low binding energy is a QWS which originates from the confinement of bulk conduction-band electrons along the thickness of the thin-film.

The spin-resolved measurements were performed with 8.5~eV photons since this photon energy provided the best trade-off between photoelectron intensity and IMFP. The measurements were done for varying emission angles while the measured kinetic energy of the emitted electrons was fixed. In this way, spin-resolved data have been acquired in the form of MDCs at the Fermi level. The reciprocal-space resolution of the experiment was about $0.03$~\AA$^{-1}$ and the energy resolution 150~meV. Figure~\ref{fig:arpes}(e) shows the experimental Fermi surface (FS) measured at a photon energy of 8.5~eV, and Fig.~\ref{fig:arpes}(f) sketches the reciprocal-space lines, with respect to the spectral features observed in the FS, along which spin resolved MDCs were acquired.

\section{Modelling the spin structure}
\begin{figure*}
\includegraphics[width=.9\textwidth]{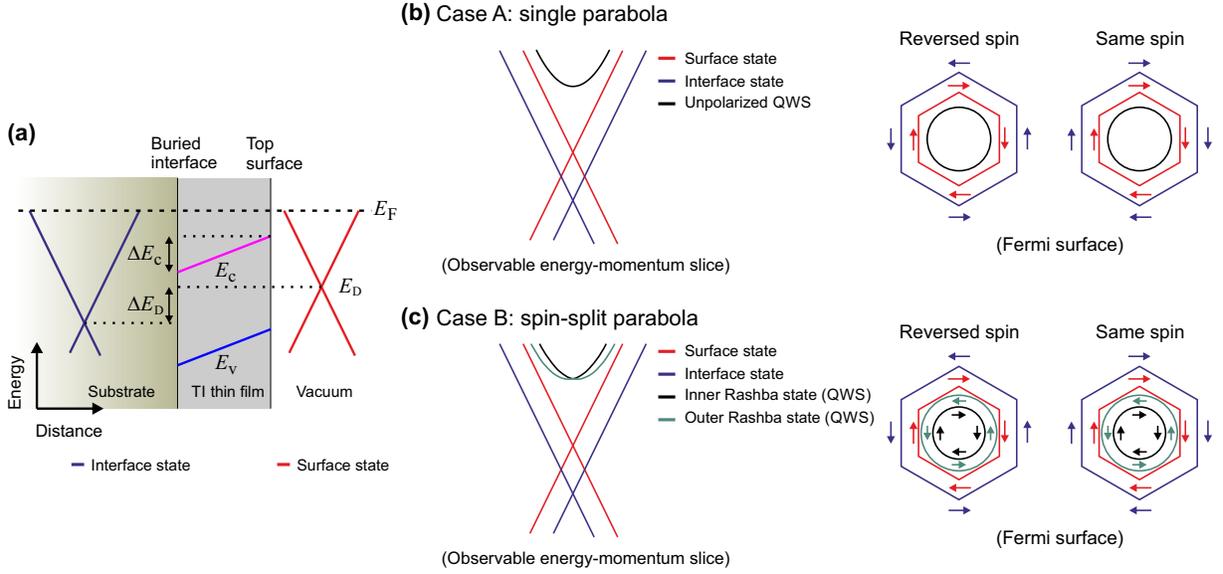}
\caption{\textbf{Model of spin structure.} (a) Schematic energy diagram of the band bending across a TI thin film. $E_\mathrm{C}$ and $E_\mathrm{V}$ refer to the conduction-band minimum and valence-band maximum, respectively, of \BiSe. The interface Dirac point ($E_\mathrm{D}$, on the left) is shifted in energy ($\Delta E_\mathrm{D}$) compared to the Dirac point of the surface state at the solid-vacuum interface. For simplicity, a linear band bending, $\Delta E_\mathrm{C}$, is assumed. (b) Two hypothetical configurations for the spin polarization of the experimentally observed bands. The double-X-shaped feature could either have parallel or anti-parallel spins while the quantized conduction band state is unpolarized. A MDC cut across the center of these features will observe a total of four polarized branches. (c) If the quantized state is also polarized, a total number of eight polarized branched are observed when looking at a MDC through these features. The polarization of the outer two double-X-shaped features can be either parallel or anti-parallel. The validity of the four cases outlined in (b) and (c) are checked by comparing the theoretical spin polarization for each case with the experimentally determined spin polarization.}
\label{fig:model}
\end{figure*}

In Fig.~\ref{fig:arpes}(f), we have already sketched the proposed electronic and spin structure of the features that can be observed in the energy-momentum and FS slices in panels (a)--(e) in the same figure. The structure originates from the fact that a TI will have topological surface states present at all interfaces towards materials that are topologically trivial. This warrants the presence of a Dirac state both at the surface (vacuum interface) and the interface towards the substrate. Provided that the interface state can be probed, i.e. the electron IMFP is sufficiently large, one will observe a photoemission spectrum that is a superposition of two Dirac states. We have previously observed \cite{Berntsen2013} that although this is the case, the Dirac points of the two states will be shifted in energy due to a band bending across the thin film, see Fig.~\ref{fig:model}(a). The states are therefore separated in energy and can be distinguished. This is in agreement with the current observations where the energy separation of the surface and interface Dirac points amounts to 380 meV, as extracted from Fig.~\ref{fig:arpes}(a). 

Without any prior experimental knowledge concerning the spin structure of the two Dirac states or the parabolic QWS seen in the ARPES spectra four possible scenarios can be envisioned, as presented in Figures~\ref{fig:model}(b) and ~\ref{fig:model}(c). The QWS are possibly Rashba-split and therefore spin-polarized, which gives rise to the general cases A and B with only a single spin-degenerate parabola or a spin-split parabola, respectively. Additionally, the two Dirac states can either have the same spin polarization (“Same spin”) or the opposite (“Reversed spin”).  

The expected total spin polarization corresponding to the four scenarios outlined in Fig.~\ref{fig:model} can be calculated and compared to the experimentally determined spin polarization. In the following, we will do this by initially fitting a number of Lorentizian peaks to the measured spin-integrated MDCs, using the peak positions and amplitudes as fitting parameters. The experimental spin-integrated MDC signal is given by the sum of the intensities ($I_i$) from the four channels of the Mott detector, i.e. 
\begin{equation}
I(k) = \sum_{i=1}^{4} I_{i}(k).    
\end{equation}
Once Lorentzian peaks have been fitted to the MDC, the spin polarization, $P(k)$, can be calculated by 
\begin{equation}
P(k) = \frac{\sum_{i=1}^{N}L_i(k)p_i}{\sum_{i=1}^{N}L_i(k)},
\label{eq:polarization}
\end{equation}
where $L_i(k)$ is the intensity of the $i$:th Lorentzian in the reciprocal-space point $k$, and $p_i$ its degree of polarization given in the interval $[-1,1]$. The sums in Eq.~(\ref{eq:polarization}) run over $N$ Lorentzians, representing the number of bands traversed by the MDC. The sign of the polarization of each Lorentzian is defined such that for a state with a spin vector rotating clockwise around $\bar{\Gamma}$, the polarization of a MDC measured from negative to positive $k$ values goes from negative to positive. Finally, the calculated polarization in Eq.~(\ref{eq:polarization}) is fitted to the experimentally determined polarization by using the individual polarizations, $p_i$, as fitting parameters.

 \section{Results}
With the current experimental setup, the in-plane and out-of-plane spin components can be determined~\cite{Berntsen2010}. The in-plane component is along a direction parallel to the entrance slit of the hemispherical analyzer which corresponds to the component in the $k_x$ direction shown in Fig.~\ref{fig:arpes}(e) and~\ref{fig:arpes}(f). In our analysis we focus only on the in-plane component since this is the dominating polarization direction of the Dirac state~\cite{Pan2011}. 

The experimental spin polarization is determined by calculating the asymmetry of the measured intensities in the spin-up ($\uparrow$) and spin-down ($\downarrow$) channels of the Mott detector divided by the Sherman function of the detector according to
\begin{equation}
P(k) = \frac{1}{S_\mathrm{eff}} \frac{I_\uparrow(k)-I_\downarrow(k)}{I_\uparrow(k) +I_\downarrow(k)},
\label{eq:effpol}
\end{equation}
where $S_\mathrm{eff} = 0.17$ is the experimentally determined effective Sherman function~\cite{Berntsen2010}.

\begin{figure}
    \centering
    \includegraphics[width=\columnwidth]{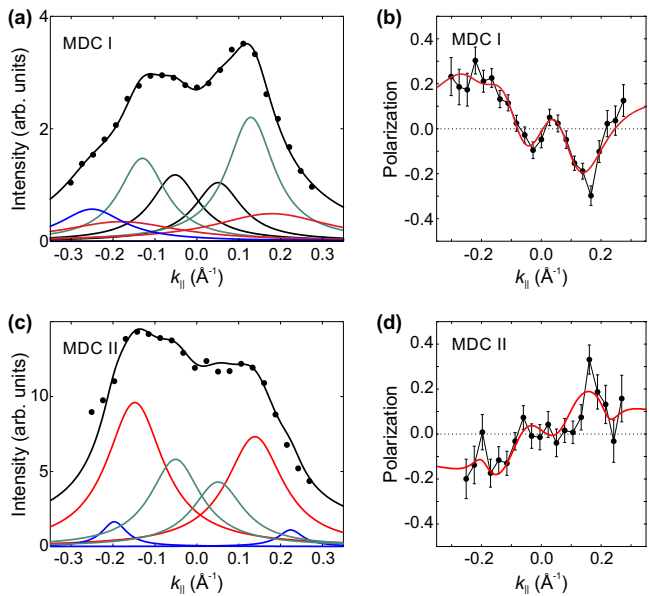}
    \caption{\textbf{Experimental spin polarization.} (a)--(b) Spin-integrated intensity for MDC~I and the resulting projected in-plane spin polarization in the direction perpendicular to the scanning direction. Solid lines are fits to the data, a total number of 8 peaks are used in the fit. The colored solid lines in (a) show the contributions of the individual peaks with the effective polarizations (from left to right) $0.7$, $-0.6$, $0.5$, $-0.5$, $0.5$, $-0.5$, $0.6$, $-0.7$. (c)--(d) Spin-integrated intensity for MDC~II and resulting projected in-plane spin polarization in the direction perpendicular to the scanning direction. Solid lines are fits to the data, a total number of 6 peaks are used in the fit. The colored solid lines in (c) show the contributions of the individual peaks with the effective polarizations (from left to right) $0.5$, $-0.35$, $0.3$, $-0.3$, $0.35$, $-0.5$. Note that the “inner part” of the Rasha-split QWS is not observed in this scan, thus only 6 peaks are used.}
    \label{fig:mdcpol}
\end{figure}

Figure~\ref{fig:mdcpol} presents the spin-integrated intensities (solid markers in panels (a) and (c)) along with the corresponding photoelectron polarizations (black solid line with solid markers in panels (b) and (d)) calculated using Eq.(\ref{eq:effpol}). The data have been measured along the two reciprocal-space lines sketched in Fig.~\ref{fig:arpes}(f). As shown in panels (a) and (d), the MDCs can be seen as a superposition of various peaks resulting from crossing the different spectral features while traversing the sketched reciprocal-space lines. 

The MDC intensities have been fitted using Lorentzian peaks given the constraints that; i) the peak positions are limited to a narrow angular range around their expected positions obtained from the ARPES data, and ii) the Lorentzian widths are pairwise locked between peaks at opposite momenta. The spin polarization, calculated using the fitted peaks and Eq.~\ref{eq:polarization}, has then been fitted to the experimental polarization assuming alternating signs for the corresponding polarizations of the individual peaks while pairwise locking the degree of polarization ($p_i$) for peaks at negative and positive momenta. This scenario thus corresponds to “Case-B with Reversed spin” as sketched in Fig.~\ref{fig:model}(c). 

The details of the analysis of MDC~I and MDC~II differ slightly. While MDC~I shows contributions from cutting in total four times through the branches of the Rashba-split QWS, MDC~II contains only contributions from the outer branch, consistent with its much larger distance to $\bar{\Gamma}$, as seen in Fig.~\ref{fig:arpes}(f). This means that MDC~I is fitted with a total number of eight peaks while six peaks are used for MDC~II. Also, as already expected from the relative intensities in the ARPES spectrum in Fig.~\ref{fig:arpes}(e), the outermost features are very weak in comparison to the QWS and the surface state. Thus, the measured polarization of MDC~I is dominated by the QWS and the contributions from the other features are difficult to identify. On the other hand, the peak fit and the measured polarization of MDC~II indicates that the surface state is the major contributor in this case. Although the outermost peaks are substantially weaker compared to the surface state peaks their influence on the total calculated polarization can be seen as a reduction of the magnitude of the polarization in the vicinity of their peak positions. Similar “dips” in the polarization are also present in the experimental data close to $k=\pm 0.2$~\AA$^{-1}$ in Fig.~\ref{fig:mdcpol}(d).

An overall lower degree of polarization is found for the spectral features in MDC~II compared to MDC~I. This is consistent with MDC~II being acquired at a larger $k_x$-value (see Fig.~\ref{fig:arpes}(f)) and hence the measured $k_x$-component of the in-plane spin is smaller than for MDC~I. 

\begin{figure}
    \centering
    \includegraphics[width=0.8\columnwidth]{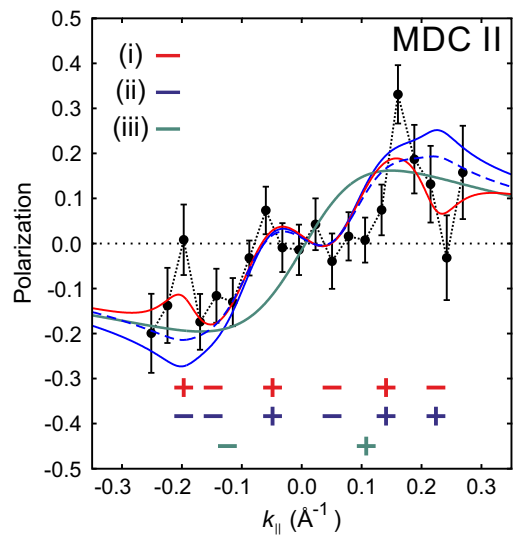}
    \caption{\textbf{Comparing different spin scenarios.} Projected experimental in-plane spin polarization for MDC~II (black dotted line with solid markers, same data as in Fig.~\ref{fig:mdcpol}(d)). Colored lines depict the expected polarizations for differently modeled theoretical scenarios: (i) Solid red line: alternating polarizations (same as in Fig.~\ref{fig:mdcpol}(d)) for a total number of six peaks; (ii) Solid blue line: same as (i), except that the sign of the polarization of the outermost peaks are reversed. Dashed blue line: refitted polarization assuming the surface-state peaks and the outer peaks carry the same polarization direction. Resulting polarizations of the individual peaks are (from left to right) $-0.3$, $-0.3$, $0.25$, $-0.25$, $0.3$, $0.3$; (iii) Solid green line: the polarization from two surface-state peaks only (polarization $\pm0.25$). Red, blue and green $-$ and $+$ signs indicate the sign of the polarization for the individual peaks in the different scenarios (i), (ii) and (iii), respectively.}
    \label{fig:polcomparison}
\end{figure}

In order to more clearly visualize the contribution of the outermost peaks to the overall polarization, Fig.~\ref{fig:polcomparison} compares the calculated polarization from Fig.~\ref{fig:mdcpol}(d) [line (i)] with the case where the sign of the polarization for the outer peaks is reversed. This artificial situation, corresponding to “Case B with Same spin” (Fig.~\ref{fig:model}(c)) where the polarization is reversed without refitting its magnitude, illustrates the extreme behavior of polarizing the interface state in the same direction as the surface state [line (ii)]. As a result, “humps” are now observed in the polarization in the opposite direction compared to the “dips” in the data. Naturally, using this particular polarization direction of the outer state, the best fit to the overall data would be achieved by reducing the degree of polarization of both the surface state and the outer peaks, as seen from the dashed (blue) line in Fig.~\ref{fig:polcomparison}. However, doing so does not change qualitatively the behavior of the polarization and the observed “dips” can never be recovered.

In Fig.~\ref{fig:polcomparison}, we also plot the polarization obtained by assuming there are only two oppositely polarized peaks in the MDC, and polarizing them in accordance with the expected surface state polarization [see line (iii)]. This curve does not describe the data well, since the MDC clearly cannot be fitted by two peaks only, but it underlines the importance of including the QWS peaks in the fit as well as assigning them a finite polarization. 

From Fig.~\ref{fig:polcomparison} it is clear that the details of the data are described best using alternating polarizations as sketched in Fig.~\ref{fig:arpes}(f). This suggests that the outer state is indeed observed in our spin measurement and that it has the opposite polarization as compared to the surface state. This is consistent with the interpretation of the observation of both the surface and interface states (as well as the Rashba-split QWS) in a 6~QL \BiSe topological-insulator film grown on Si(111).

\section{Discussion}
The surface state in the three-dimensional TIs, e.g. \BiSe and \BiTe, is expected to display a spin-momentum locking with a high degree of spin polarization and a polarization vector that lies predominantly in the surface plane. In spite of first principles calculations suggesting a significant decrease in polarization due to strong spin-orbit entanglement \cite{Yazyev2010}, previous spin-resolved ARPES studies have demonstrated polarizations ranging from 60~\% to nearly 100~\%~\cite{Souma2011,Pan2011,Jozwiak2011,Hsieh2009}. For \BiSe and \BiTe samples that are highly electron doped the Fermi surface of the Dirac state becomes hexagonally warped~\cite{Kuroda2010} and an out-of-plane spin component has been observed in \BiTe~\cite{Souma2011}. 

In the present study, the polarization determined from the in-plane channels of the Mott detector is expected to be large due to the sensitivity of the spin component perpendicular to the reciprocal space lines drawn in Fig.~\ref{fig:arpes}(f). However, due to the clear haxagonal distortion of the Fermi surface (see Fig.~\ref{fig:arpes}(d)--(e)) a reduction of the in-plane polarization is anticipated. Scattering of the helical Dirac fermions with un-polarized bulk or defect states could also lead to a reduced polarization in the case when the Fermi level is positioned in the conduction band~\cite{Park2010}.

The effective polarizations found for the peaks in MDC~I (from left to right in Fig.~\ref{fig:mdcpol}(a): $0.7$, $-0.6$, $0.5$, $-0.5$, $0.5$, $-0.5$, $0.6$, $-0.7$) suggest a high degree of polarization for the probed states. As MDC~I is taken close to $k_x = 0$~\AA$^{-1}$ we expect the in-plane spin to be parallel/anti-parallel to the $k_x$-direction, with a minimal component along $k_y$, and we therefore take the measured polarization as the magnitude of the in-plane spin. Figure~\ref{fig:arpes}(f) shows that MDC~I in fact was not acquired at a constant $k_x$-value which means that our assumption of a negligible $k_y$-component of the spin is strictly not valid. However, due to the small deviation from a perfectly vertical MDC, see Fig.~\ref{fig:arpes}(f) ,we expect our assumption to give a fair approximation of the magnitude of the in-plane spin.

Figure~\ref{fig:arpes}(d)--(f) suggest that the FS of the surface and interface states are hexagonally warped. Based on the different sizes of the two FS, and assuming a perfect hexagonal behaviour, we estimate that at the position where MDC~II (acquired at $k_x \sim 0.2$~\AA$^{-1}$) intersects the outer hexagon (interface state) the in-plane spin has an angle of $\alpha_\mathrm{outer}=30^{\circ}$ with respect to the $k_x$-component that we are measuring. On the other hand, the inner hexagon (surface state) is intersected close to a corner where the exact angle is more difficult to estimate. However, taking into consideration that our peak fitting results in two peaks for the inner hexagon as well, we can assume that the corresponding angle for the in-plane spin at this position of MDC~II is larger than 30$^{\circ}$ but significantly smaller than 90$^{\circ}$. We therefore set this angle to be $\alpha_\mathrm{inner}=45^\circ$. 

For the outer branch of the QWS, we use the $k_x$ and $k_y$ coordinates of the corresponding peaks in MDC~II to estimate the angle ($\alpha_\mathrm{QWS}$) between the in-plane spin vector and the measured projection along the $k_x$-direction -- assuming a parabolic QWS with a tangential in-plane spin vector. Using these angles, the the resulting magnitude of the in-plane spin for the different peaks in MDC~II become $0.58$, $-0.50$, $1.0$, $-1.0$, $0.50$, $-0.58$, which for the surface and interface states are closer to the values obtained from MDC~I. If $\alpha_\mathrm{inner}$ and $\alpha_\mathrm{outer}$ are increased to 55$^\circ$ and 45$^\circ$, respectively, the polarization of the surface and interface state reaches the same values as obtained from MDC~I. We interpret this as evidence for the in-plane spin to have a larger $k_y$-component at the positions measured by MDC~II -- which is logical considering that the true FS deviates from a perfect hexagon (see Fig.\ref{fig:arpes}(d)), having rounded rather than sharp corners. 

The close to full polarization of the QWS obtained by our calculations, using the measured polarizations from MDC~II, is expected for a Rashba-split QWS -- yet larger than the values obtained from MDC~I. The deviation could partially be a result of the aforementioned underestimation of the in-plane spin due to a non-zero $k_y$-component in MDC~I, but could also come from uncertainty in the exact $k_x$-value of MDC~II. To illustrate the latter, by assuming $k_x=0.16$~\AA$^{-1}$ for MDC~II, the calculated magnitude of the in-plane spin for the QWS is reduced to $\sim0.85$.

Clearly, there is some uncertainty connected to our estimates of the magnitude of the in-plane spin using MDC~II and the approach described above. Nevertheless, the resulting polarizations obtained from MDC~I and MDC~II are consistent with a model containing spin-polarized surface, QWS and interface states, respectively, with high degrees of polarization. The ability to accurately ascribe features observed in the angle-resolved spectra to either the surface or interface therefore becomes possible using the additional information contained in the spin-resolved data. 

Although the observation of surface and interface states made here are in line with the expected electronic states at the boundaries of a TI, our experiments nevertheless unequivocally establish the unchanged spin structure of the Dirac surface state in a TI when interfaced with a topologically trivial material, as well as the spin-polarized nature of the QWS in the ultra-thin limit of a TI-film. As any implementation of TI materials in actual devices would involve interfaces our results, showing a high degree of spin polarization for the interface state, are of importance to confirm the feasibility of applications that rely on the spin texture of a TI-interface state to be similar to the vacuum-surface state.

Since the spin structure of the surface state on \BiSe is know from previous studies~\cite{Jozwiak2011} our conclusions regarding the spin polarization of the TI-interface state is expected and does not provide any dramatic new insights on the spin nature of this material. However, by using these results from a rather well-known system we are able to give compelling evidence for the usefulness of SARPES conducted in the low photon-energy range in accessing spin-resolved electronic structure information from buried electronic states. Our work also underlines the importance of combining this type of experiments with a theoretical model in order to extract the relevant information from an, at a first glance, unexpectedly complex measured spin polarization. Although the theoretical model used here describes only in simple terms, qualitatively, the expected spin structure in the \BiSe thin film, it nevertheless provides a crucial addition to the analysis that allows us to decompose the total measured spin polarization into contributions from the surface, bulk and interface, respectively.

As the electronic structure at junctions between materials are subject to, e.g., band alignment (between semiconductors), charge accumulation or depletion layers or proximity effects, their configuration can dramatically differ from the intrinsic bulk properties. \textit{Interface physics} becomes an important route towards generating new electronic states that can be exploited, e.g., in device applications and motivates why characterization of the electronic and spin structure of such states is becoming increasingly important. Although our analysis presented herein is qualitative, we expect that quantitative spin information can be achieved in other cases by incorporating more advanced models and band structure calculations.

Presently, there is an ongoing interest in quantum materials where strong correlations are merged with topological matter into applications that use and control individual quantum states for quantum based computing or sensing. There is a growing interest from these device communities in using ARPES to gain electronic-structure information on their devices. Several novel spin-based components, such as the spin field-effect transistor~\cite{Datta1990,Ingla2021}  and the spin valve~\cite{Dieny1991,Bordoloi2020} are being heavily investigated, which directly rely on spin-polarized electronic states. Spin-ARPES can prove to be an important experimental tool in this field of research for gaining direct information on the spin structures that form the basis for the functional properties of such devices. 

Furthermore, the experimental setup used in this study relies on Mott-scattering as the mechanism to distinguish between different spin directions of the photoelectrons. In spite of being a robust and proven technique, the process suffers from a low efficiency and thus requires long acquisition times to achieve a sufficient signal to noise ratio. Spin polarimetry based on spin-polarized low-energy electron diffraction (SPLEED)~\cite{Kirschner1979,Yu2007} and very-low energy electron diffraction (VLEED)~\cite{Okuda2008} have demonstrated improved detection efficiency over Mott scattering and several implementations~\cite{Kutnyakhov2013,Ji2016,Okuda2015} reach a high figure of merit through parallel detection schemes. Recent detector developments using spin-dependent transmission through thin ferromagnetic layers as filtering mechanism~\cite{Overgaard2017,Tereshchenko2021} have the potential to simplify implementation of spin detectors in ARPES setups. A number of SARPES end-stations are currently under operation or commissioning at low-photon energy synchrotron beamlines~\cite{sarpesBeamlines}, which sets the scene for an increased number of device-related SARPES studies in the years to come. For the device communities, future developments combining SARPES with a nano-focus will be particularly interesting to follow.

In summary, spin-resolved ARPES (SARPES) measurements on a TI thin film are used to demonstrate the capability of this technique to gain information on the spin structure of electronic states located at a buried interface. We show that a careful analysis, combined with a theoretical model of the expected spin structure, enables the contributions to the overall measured spin polarization from electronic states present at the surface, bulk and interface to be disentangled. Access to electronic-structure information from the buried interface and bulk is in the present study warranted by the large electron IMFP as a result of the low photon energy used in the photoemission experiment. 

We envision that the SARPES technique in the low photon-energy range can play an important role in determining the electronic spin structure with high energy resolution in electronic device architectures relevant for future quantum technologies.

\section{Acknowledgements}
This work has received financial support from the Swedish Research Council (2019-03486),  and the Knut and Alice Wallenberg Foundation (2018.0104).

\section{Author contributions}
M.H.B. and O.T. planned the experiment, all authors contributed to the experimental work, M.H.B and B.M.W performed the data analysis, M.H.B wrote the manuscript with input and comments from the remaining authors.


%

\end{document}